# Regenerative Soot: A cusp field, graphite hollow cathode, carbon cluster ion source

**Shoaib Ahmad[1,2,*] and Tasneem Riffat[1]**

[1]*Accelerator Laboratory, PINSTECH, P.O. Nilore, Islamabad, Pakistan*
[2]*National Centre for Physics, Quaid-i-Azam University Campus, Shahdara Valley, Islamabad, 44000, Pakistan*

Email: sahmad.ncp@gmail.com

## Abstract

An all carbon hollow cathode ion source with a specially designed cusp magnetic field $B_z(r,\theta)$ is described. Velocity spectra from this source has shown that carbon cluster synthesis is taking place as a function of the source design, discharge parameters and the state of sooting. The cusp magnetic field seems to establish two distinct discharge regions within the source. The first is a narrow annular ring that is 2 mm thick and 15-20 mm long with restricted axial and radial extent of the cusp field. Here the source gas ionizes. The carbon atoms sputter of the cathode walls and subsequently get ionized. The other one is a cylindrical plasma region with the entire axial range of $B_z(r,\theta)$. In this region the dynamics of the soot formation from the plasma constituent on the cathode walls seems to be related to the carbon clustering processes

## 1. Introduction

Carbon clusters $C_m$ (m > 1) can be formed by various methods [1-9]. The common feature of these methods involve generating of the C bond breaking and re-bonding sequences. All of these relate to the carbon bonded with its nearest neighbors in the condensed media. We have investigated the dynamics of soot formation from sputtered C atoms and ions in magnetized plasma to produce a range of $C_m$ for cluster - solid interaction studies. Cold cathode uno- and duoplasmatron sources [10-12] with hollow cathodes operating in Penning discharge mode with axial or radial magnetic fields can yield high cur-rent densities of singly and multiply charged gaseous ions. In our present design of the hollow cathode cluster ion source we have utilized the phenomenon of the cathode wall sputtering by the support gas ions. The wall sputtering has been used as a means to introduce C into the plasma. The glow plasma discharge at low pressures ~$10^{-1}$-$10^{-3}$ mbar and discharge voltages 0.5 -1.0 kV with cusp magnetic field provides the environment in which the C atoms ionize, form soot layers on cathode walls and initiate cluster synthesis processes.

Operation of the source consisting of a cylindrical graphite tube- the Hollow Cathode (HC) and a coaxial Hollow Anode (HA) with a specially designed multi-cusp magnetic field depends on HC sputtering efficiency and soot forming properties leading to the clustering of carbon atoms and ions. A steady stream of carbon atoms is sputtered into the glow discharge plasma from graphite HC surface. The key to the ignition and sustenance of the discharge at pressures $\sim 10^{-1}$-$10^{-3}$ mbar is a set of six bar magnets wrapped around HC providing an axially extended set of cusp magnetic field contours. The hexapole field confinement is designed so that the r, $\theta$ and z components of the field lines produce the combined magnetic field contours $B_z(r,\theta)$ extending over the entire HC region. Whereas, molecular gases $CO_2$, $CH_4$, $N_2$, noble gases i.e., Ar and Ne and mixtures Ar + $N_2$ and Ne + $N_2$ have been used, Ne has proved to be the most efficient support gas to provide sooting plasma. It may be due to the proximity with car-bon in mass and also in its lowest excitation potential ≈16:6 eV being higher than the ionization potentials of carbon and most of its clusters. The sooting plasma so produced demonstrates a temporal growth in the densities of sputtered carbon atoms and ions as a function of the discharge voltage $V_d$ and current $i_d$. Once $C^+_1$ ions anchor onto one of the field contours, the direction of their consequent gyrational motion and clustering probability is determined by collisions with electrons, neutral and excited C and the support gas atoms. The streams of gyrating $C^+_1$, $Ne^+$ and $C^+_m$ (m ≥ 2) ions with large collision cross sections eventually lead to the inside walls of HC where they impact with E ≈ $qV_d$. The impact continuously modifies the cathode wall's material characteristics and covers it with the gradually increasing layers of soot. The soot subsequently becomes the surface for later sputtering to take place. Velocity spectra of emitted charged species have shown the dependence of carbon cluster emission on the state of sooting in the source.

## 2. Experimental

The design features and operational characteristics are shown in Fig. 1. The HC and a partially penetrating HA is shown in Fig. 1(a) with the relative position of the set of six bar magnets. Control of the intensity and profile of the $B_z(r,\theta)$ field in the annular as well as open cylindrical re-gions between HC and HA is achieved by mild steel (MS) rings and plates. A MS base plate holds the source composed of an alumina tube bonded at both ends with MS rings that hold the respective electrodes HC and HA. The hollow anode penetrates the cathodic cylinder up to the radial plane where $B_z(r,\theta)$ has a maximum as a function of z as shown in Fig. 1(c). The bar magnet set is held on the inner surface of an MS ring that acts as a field return core, magnet support and heat sink. Fig. 1(b) shows the cross section of the cusp field lines that are intersected by the hollow cathode at $r_{max} = r_{HC}$ and shown as black circle while a dotted circle indicates the outer radius $r_{HA}$ of the hollow anode. The axial variation of $B_z(r,\theta)$ for different values of r as a function of z is presented in Fig. 1(c) where the flux density can be seen as maximum at the centre of bar magnets and the hollow anode extends to this plane. A cylindrical MS ring screwed over HA shapes the 3-D $B_z(r,\theta)$ as a function of z cusp field configurations, which are seen to be crucial to the source's operational as well as clustering characteristics.

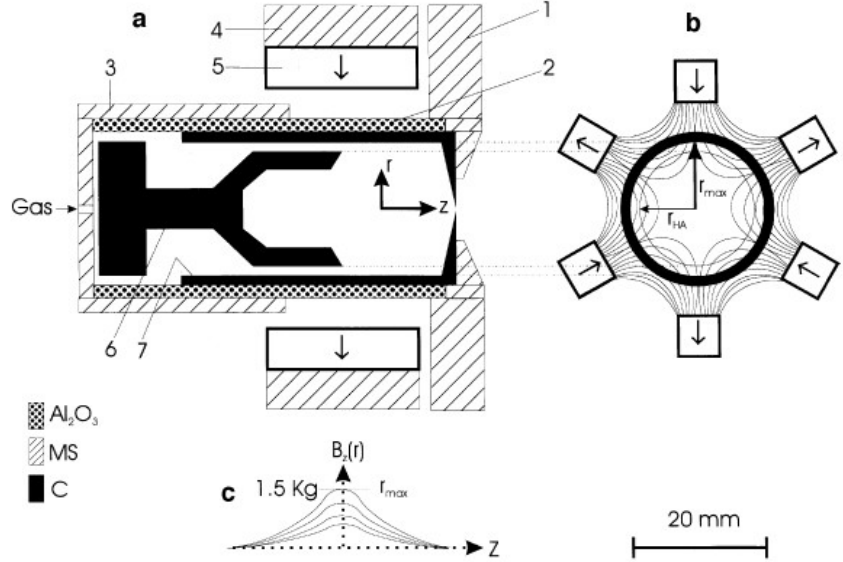

Fig. 1. The source is shown in (a) with the HC and HA and a set of six permanent bar magnets. A base plate (1) holds an lumina tube (2) bonded at both ends with MS rings holding HC shown as 6 and HA as 7, respectively. The bar magnet set (5) are held on the inner surface of an MS ring (4). (b) shows the r and $\theta$ variations of cusp field lines intersected by HC shown as a thick black circle at $r_{HC} = r_{max}$ and a dotted circle indicates HAs outer radius $=r_{HA}$. The axial field $B_z(r,\theta)$ for a given h, is presented in (c). Cylindrical MS ring (3) shapes the desired $B_z(r,\theta)$ vs. z cusp field contours.

The cusp magnetic field [13] has been incorporated into different structural designs of the hollow cathode source in such a way that it helps to produce two distinct discharge regions for all configurations; the first is the narrow annular region between the outer walls of HA and the inside of HC extending from the tip of HA backwards. This is a thin 2 mm thick disc of 15 mm length within $B_z^{max}(r,\theta) \sim 0.1 \pm 0.15$ T. The cusps and edges produce three major elliptic zones on the outside of HA due to the impact of electrons and three regions on the inside of HC. The HC elliptic zones have enhanced sputtering activity due to the field directed $Ne^+$ and $C^+_m$ ($m \geq 1$) impact. The cathodic elliptic areas are the regions of electron emission, high plasma-wall collision activity, removal of wall material and possibly, the ion induced clustering in the sooted layers. Another cylindrical region of much larger dimension is outside the HA up to the HCs cylindrical region with the $r_{HC} = r_{max}$ and of length 20 mm 5 mm. It has the complete range of cusps and edges of the $B_z(r,\theta)$ field. The larger variation in $B_z(r,\theta)$ along with plasma filling the entire region indicates inherent clustering environment where a broad range of collisions between the glow discharge plasma species are possible.

Mass analysis of the extracted plasma species is performed with a permanent magnet based velocity filter. Velocity analysis has certain advantages over competitive mass analyzing techniques. Ideally, a velocity spectrum contains all velocities from $v_{min}(\equiv 0)$ to $v_{max}$ corresponding to masses $m_{max}(\equiv \infty)$ to $m_{min}$. In practice, higher extraction voltages help to spread out the heavier masses. Cluster identification with 2 amu for the heavier fullerenes is achieved with the velocity analyzer's magnetic field within easily attainable limits 0.4 T on the axis. Fig. 2 shows the complete experimental set-up. Cluster source is shown in the cathodic extraction mode. The plasma species are extracted from 0.6 to 0.8 mm hole with a 2 mm diameter extraction aperture. A set of three accelerating electrodes is used for the extraction and focusing of the plasma species. The role of these electrodes is crucial to yield pA beams through the beam divergence limiting collimators that are used for 1 beam into the velocity filter.

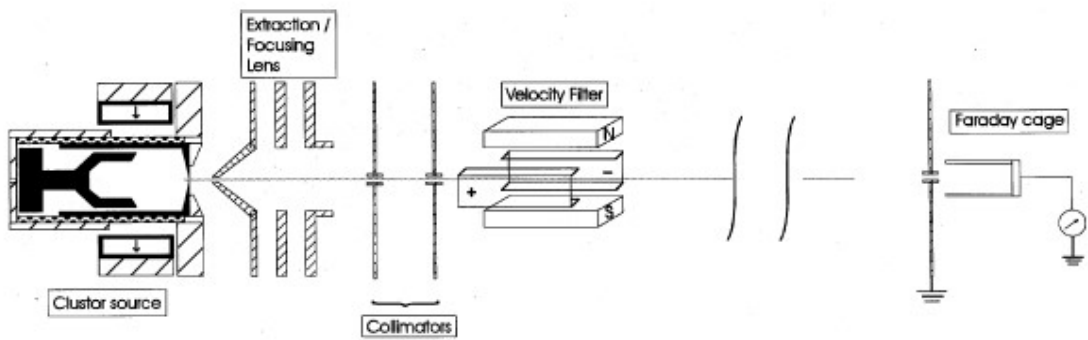

Fig. 2. The complete experimental set-up is shown with the cluster source in cathodic extraction mode. A set of three accelerating electrodes is used for the extraction and focusing of the plasma species. The velocity filter utilizes a set of four permanent magnets with a maximum field $\approx 0.35$ T on the axis. Two C shaped aluminium electrodes provide the electric field as well as hold the insulated magnets. A 1500 mm flight path provides adequate resolution to the analyzed beam that is monitored by a Faraday cage.

The permanent magnet based velocity filter is designed for a variable resolution so that the heavier cluster identification can be improved. It utilizes a set of 4 magnets with a maximum field of 0.35 T on the axis. Two C-shaped aluminium electrodes provide the electric field as well as hold the insulated magnets. The edge effects are compensated by ex-tending the electrodes and using MS field shortening discs. A 1500 mm flight path provides adequate resolution to the analyzed beam that is monitored by a Faraday cage. Data acquisition is done by a PC controlled electrostatic dual power supply of the velocity filter and cluster ion current is measured by a pA meter. For even higher resolution larger flight paths with Channeltron detector can be employed. The dispersion D of masses $m_0 \pm \delta m_0$ from the straight through resolved mass $m_0$ is given by [14,15] as $D \propto al (\delta m_0/m_0)(\varepsilon_0/V_{ext})$, where $a$ and $l$ are the lengths of the velocity filter and the flight path, respectively. The choice of the extraction voltage $V_{ext}$ for a fixed magnetic field $B_0$ determines the particle velocity $v_0 (=\varepsilon_0/B_0)$ which in turn fixes the compensating electric field $\varepsilon_0$. Manipulation of $a$ and $l$ help in varying the dispersion according to the experimental requirements. The optimum value of $l$ depends on the mass resolved beam diameter $b$ at the detector such that $D \geq b$. The filter's axial dimension a is a design parameter and relatively more difficult to increase compared with $l$. How-ever, multiple units of the velocity filter can be stacked to increase D as $D \propto na$, $n$ being the number of such units.

# 3. Results

## 3.1. The role of hollow cathode sputtering

The state of hollow cathode's inner walls due to adsorption is an essential feature for determining the source's operational characteristics. These include electron emission, sputtering of HC leading to the inclusion of HC material into the plasma and the gradual build up of the adsorbed layers. A competition between the dynamics of HC wall erosion and the adsorbed thin film seems to determine the state of clustering. The most important parameter being the level and the type of sooting. In this section we present results from the source operated with those gases or mixtures of gases that are effective sputtering agents but operate the source in an un-sooted mode.

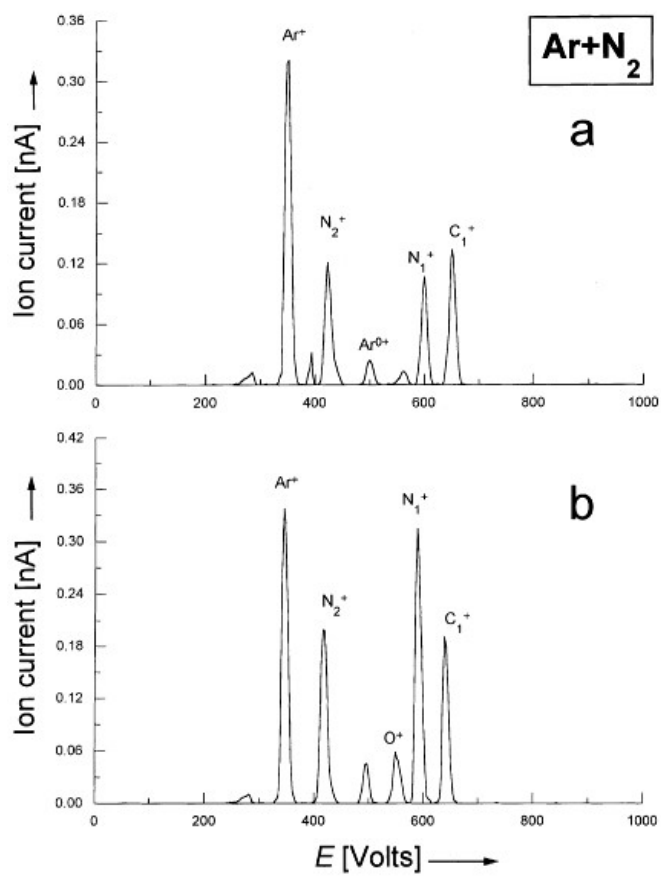

Fig. 3. Ar + 1% $N_2$ mixture is used and the two spectra are presented that show the relative contribution of $Ar^+$, $N_2^+$ and $N_1^+$ along with the $C_1^+$ peak. (a) was obtained at $i_d$ = 100 mA. In (b) the $N_1^+$ peak shows an increase at a lower discharge current $i_d$ =50 mA.

The first series of experiments was conducted with molecular gases containing carbon i.e. $CO_2$ and $CH_4$ to produce carbonaceous plasma. The plasma so produced did yield a high percentage of $C_1^+$ that depended on $i_d$ and other relevant gas discharge parameters. The next batch of experiments was with mixtures of Ar and Ne with 1±5% $N_2$. Experiments were also done with pure $N_2$. Though the heavier noble gases like Xe, Kr and Ar have lower ionization potentials and higher sputtering yields but have proved to be difficult to ionize in our particular source geometry. On the other hand molecular gases and their mixtures with noble gases are relatively easy to initiate the discharge. All of these experiments relied on the fact that the sputtering of HC introduces $C_1$ into the plasma which in turn can be ionized and be-come an active plasma constituent. $C_1^+$ has been seen in the velocity spectra from the above mentioned support gas plasmas.

Enhanced HC sputtering that in turn intro-duces C into the plasma is also a hallmark of Ar and Ne mixtures with $N_2$. In Fig. 3 Ar + 1% $N_2$ mixture is used and the two spectra are presented that show the relative contribution of $Ar^+$; $N_2^+$ and $N_1^+$ along with the $C_1^+$ peak. Fig. 3(a) was obtained at $i_d$ = 100 mA. The $N_1^+$ peak shows an increase in Fig. 3(b) at a lower discharge current $i_d$ = 50 mA. However, the overall pattern of the two spectra in Figs. 3(a) and (b) is same except that the ionization efficiency of the plasma constituents varies. There is no evidence of the $C_2^+$ in Ar + 1% $N_2$ mixture.

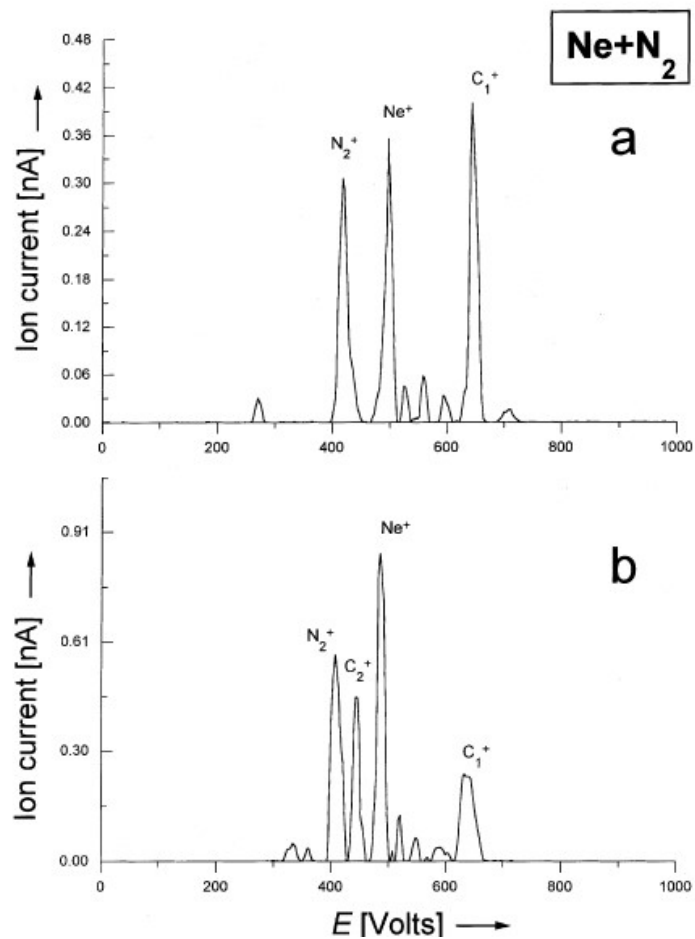

Fig. 4. In (a) $N^+$, $Ne^+$ and $C_1^+$ peaks are the major participants of the plasma in a discharge at $V_d$ =0.6 kV, $i_d$ =200 mA and the source gas pressure~ $10^{-2}$ mbar. The spectra in (b) was obtained after ~100 min operation at $i_d$ =200 mA. The emergence of the $C_2^+$ peak is the main feature along with the broad $C_1^+$ peak.

Ne + 1% $N_2$ mixture on the other hand shows a different pattern in the discharge constituents' variation as a function of $i_d$ and the resultant growth of $C_2^+$ peak. A set of two representative spectra is shown in Fig. 4(a) and (b). In Fig. 4(a) $N_2^+$, $Ne^+$ and $C_1^+$ peaks are the major participants of the plasma in a discharge at $V_d$ = 0.6 kV, $i_d$ =200 mA and the source gas pressure $10^{-2}$ mbar. Fig. 4(a) was obtained after 30 min operation at $i_d$ = 200 mA. The emergence of the $C_2^+$ peak in Fig. 4(b) has its relevance to the clustering environment in the source. The peak height of $C_2^+$ depends on the discharge conditions and shows a consequent increase with $i_d$.

### 3.2. Formation of $N_3$ and $C_2$: the onset of clustering

Operation of the source with $N_2$ establishes stable plasma that produces the usual $N_2 \rightarrow N_2^+ + e$ and the by-products via $N_2^+ \rightarrow N_1 + N_1^+$. In addition, the sputtered C and the dissociated N become the active ingredients. Physically, the ionized C and N become active participants in the plasma - HC wall interactions via sputtering and secondary electron emission mechanisms. Chemically, reactions like $C_1 + C_1 \rightarrow C_2$ and $N_1 + N_2 \rightarrow N_3$ initiate molecular synthesis processes.

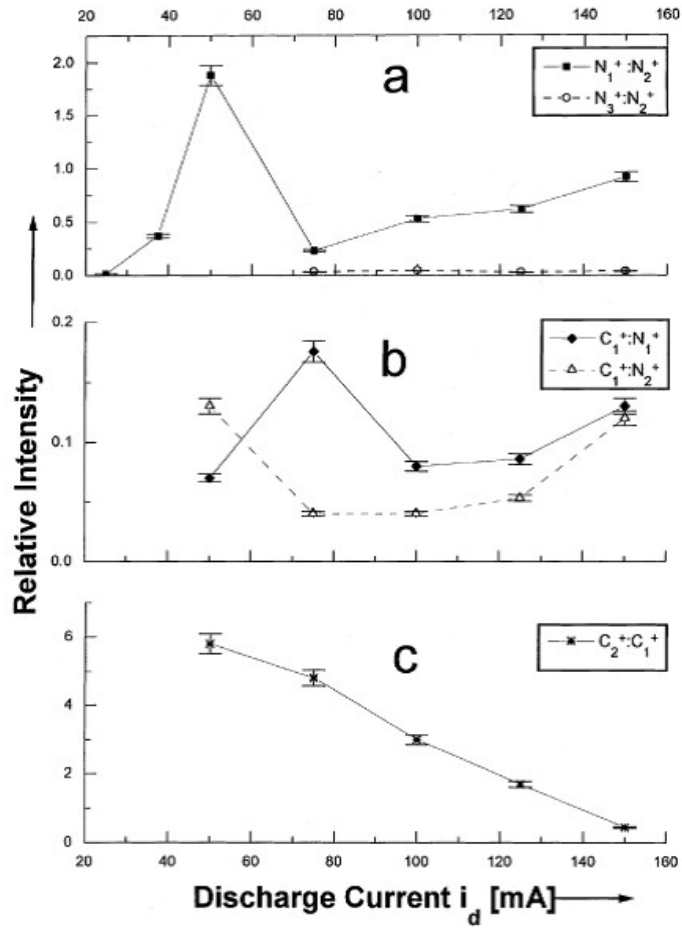

Fig. 5. In (a) the relative intensities of $N_1^+ : N_2^+$ and $N_3^+ : N_2^+$ have been plotted against $i_d$. Higher intensities of $N_1^+$ correspond to larger values of $N_3^+$: $N_2^+$ can be seen as a smaller constituent with intensity decreasing with $i_d$ in the range 75-150 mA. (b) shows ratio of the intensities of $C_1^+ : N_1^+$ and $C_1^+ : N_2^+$. A reduced sputtering due to $N_2^+$ bombardment compared with that due to $N_1^+$ is evident from (b). (c) presents the ratio $C_2^+ : C_1^+$ that reduces as the discharge power ($\propto i_d$) is increased.

We have analyzed the data from $N_2$ discharge to evaluate the relative intensity ratios of $N_3$ and $C_2$. In Fig. 5(a) the relative intensities of $N_1^+ : N_2^+$ and $N_3^+ : N_2^+$ have been plotted as a function of the discharge current $i_d$. The discharge current $i_d$ is controlled with a variable voltage and constant current power supply (2.5 kV, 200 mA). Higher intensities of $N_1^+$ yield correspondingly larger values of $N_3$. $N_2$ dissociation and ionization goes through a peak around 50 mA and gradually equilibrates with $N_1$ at higher $i_d$. $N_3^+$ can be seen as a smaller constituent of the plasma and its intensity gradually decreases with $i_d$ in the range 75-150 mA. Fig. 5(b) shows ratio of the intensities of $C_1^+:N_1^+$ and $C_1^+:N_2^+$. The data seems to suggest that $N_1^+$ is more efficient in sputter/desorbing $C_1$ from HC walls compared with $N_2^+$. The bombarding energies of $N_1^+$ and $N_2^+$ ions from the plasma through the cathode sheath is between 0.5-0.7 keV. Energy of the impinging $N_2^+$ on HC surface may be consumed partly in dissociation and distribution among the $N_1$ constituents. A reduced sputtering due to $N_2^+$ bombardment compared with that due to $N_1^+$ is evident from Fig. 5(b). Fig. 5(c) presents the ratio $C_2^+:C_1^+$. The synthesis of $C_2$ goes through maximum $\approx$ 75 mA and reduces as the discharge power $\propto i_d$ is increased. Higher $i_d$ seems to imply dissociation of $C_2$ in an $N_2$ initiated plasma.

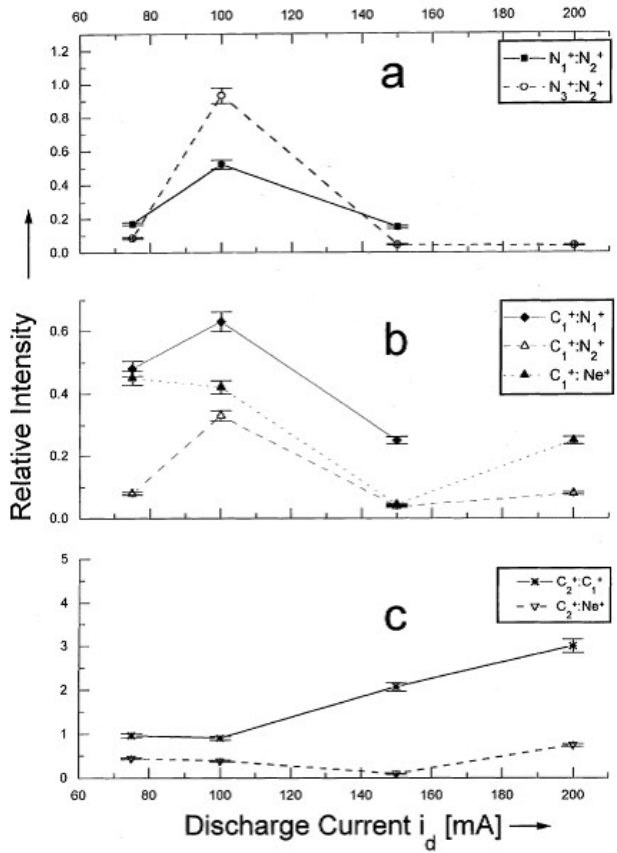

Fig. 6. Results from Ne + 1% $N_2$ mixture show some differences compared with those in Fig. 5. (a) ratios of $C_1^+:N_1^+$, $C_1^+:N_2^+$ and $C_1^+:Ne^+$ are plotted as function of $i_d$. The $C_1^+$ contribution is maximized around 100 mA in comparison with all other ion species. In (b) the $N_3^+$ intensity is much more enhanced and goes through a significant yield in this discharge current range. (c) has two sets of data plotted, one is the $C_2^+:C_1^+$ peak area ratios and the other is that of $C_2^+:Ne^+$. There is an increasing trend of $C_2^+$ with the discharge power.

Ne + 1% $N_2$ mixture has interesting and different results for the formation of $C_2$ and $N_3$ compared with those from the source ignited on pure $N_2$. Neon with a high ionization potential (21.56 eV) is generally a difficult gas to be ionized in most sources. Its mixture with $N_2$ leads to stable source operation but with reduced $Ne^+$ ion intensities. In Fig. 6(a) ratios of relative intensities of $C_1^+:N_1^+$, $C_1^+:N_2^+$ and $C_1^+:Ne^+$ are plotted as function of $i_d$. The $C_1^+$ contribution of the plasma is maximized around 100 mA in comparison with all other ion species. Similar results were obtained from pure $N_2$ operated source, especially, the ratio $N_1^+:N_2^+$ has a similar pattern in Fig. 5(b) and Fig. 6(b). The ratio $N_3^+:N_2^+$ shows a markedly different behavior for the latter figure between 50±150 mA. The $N_3^+$ intensity is much more enhanced and goes through a significant yield in this discharge current range.

Fig. 6(c) has two sets of data plotted, one is the $C_2^+:C_1^+$ peak area ratios and the other is that of $C_2^+:Ne^+$. We see the increasing trend of $C_2^+$ peaks with increase in discharge power which in turn is related to higher ionization efficiencies for $Ne^+$. Unlike the results presented in Fig. 5(c) where $C_2^+:C_1^+$ ratio showed a continuous decrease as a function of increasing $i_d$ from an $N_2$ ignited plasma, the $Ne + N_2$ mixture provides a clustering environment in which the formation of $C_2$ is favored with higher power inputs and the $N_3$ molecule/cluster formation has a favorable formative environment as well.

### 3.3. Production of sooting plasma

Source operation on pure neon gas initially requires large power inputs 100-150 W at high gas pressures $10^{-1}$ mbar for the initiation of the discharge. Discharge power $\sim V_d i_d$ can be substantially reduced once the density of the sputtered C component of the plasma reaches a minimum level, beyond which stable source operation is achievable. We have observed that neon seems to be an efficient support gas to provide conditions leading to sooting plasma. A sooting plasma implies an environment that is conducive to continuous HC wall sputtering by $Ne^+$ and other plasma constituents like $C_1^+$ and higher carbon clusters $C_m^+$ ($m \geq 2$). This in turn leads to the wall coverage with neutral, excited and ionized $C_m^{0,*,+}$ species. The competition between wall erosion and the soot formation determines the plasma properties.

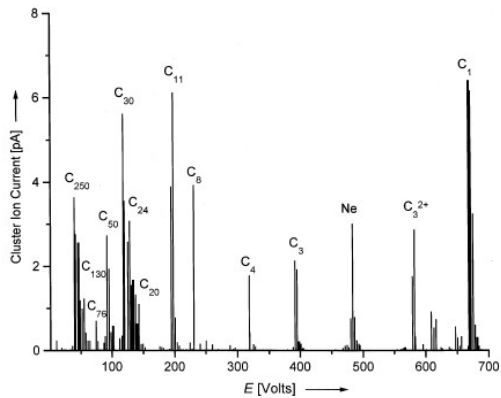

Fig. 7. It was obtained after 3 h of operation with $V_d = 0.85$ kV, $i_d = 100$ mA. A $V_{ext} = 1$ kV spectrum is composed of linear chains, rings and fullerenes. The multiply charged species in the higher velocity section of the spectra can also be seen, one such cluster shown is $C_3^{2+}$.

During the initial sooting stages of source operation $Ne^+$ and $C_1^+$ dominate the velocity spectra. By monitoring the velocity spectra one can see the temporal growth of higher clusters at given discharge conditions. Fig. 7 was obtained after 200 min with $V_d =0:85$ kV, $i_d =100$ mA and $V_{ext} =1$ kV. The spectrum has $C_1$, $Ne^+$ peaks along with clusters composed of linear chains, rings and fullerenes. Three groups of clusters $C_m$ can be distinguished in the spectrum starting from the very low velocity end i.e., lower deflection voltages for the E×B filter. The first batch is of the heavy fullerenes from m 250 to 100, thereafter follows the more familiar cluster types from $C_{76}$ to $C_{20}$ with peaks around $C_{50}$ and $C_{30}$.

The third regime of clusters ranging from m < 20 to $C_1$ are dispersed over most of the velocity spectrum.

## 4. Discussion

Various techniques exist for creating the conditions that lead to the formation of carbon clusters including chains, rings, fullerenes, onions and nanotubes [1-9]. Understanding of the mechanisms of clustering may lie in the synthesis of common features of these widely different physical methods of producing clusters. Generation of carbon plasma with laser ablated graphite followed by supersonic expansion provides an environment where normally the entire range of carbon clusters $C_m$ with $1 \leq m < 10^3$ [1,2] are produced. Tuning of the experimental conditions provides preferentially higher yields of $C_{60}$ [2]. The technique of arc discharge between the graphite electrodes produces C plasma which can, in the presence of high pressures of He or Ne, lead to clustering and large quantities of $C_{60}$ and $C_{70}$ [3].

Soot containing clusters $C_m$ with a range of m may have been produced in these experiments but only $C_{60}$ and $C_{70}$ are recovered due to their solubility in benzene and toluene. Third category is the energetic electron and ion induced clustering in the condensed media. Graphite and various carbon containing polymers have been irradiated. Microscopy of electron irradiated graphite have shown onions [4] and nanotubes [5] formation. Heavy ion irradiation with MeV [6,7] and hundreds of keV [8,9] beams has proved to provide clustering within the solid. Despite variations in these widely different energy deposition techniques, there remain similarities in the boiling of and the condensing of the C vapor, atoms and ions by either the buffer gas or the surrounding condensed media.

According to the Ecker's theory [16] our HC glow discharge source operates at low pressures with high E/p values~$10^3 - 10^6$ [V/(cm m bar)]. This implies high mobility plasma which seems to be crucial for the efficient wall sputtering in our case. Lower operating pressures reduce collision rate. Taken together, these two effects lead to the dynamical build up of sooting layers on HC providing a high enthalpy 2-D region where bond rearrangement sequences lead to cluster formation.

Hollow cathode discharge in cusp hexapole magnetic field is initiated by:

(a) the ionization of the support gas, and

(b) providing a continuous stream of sputtered carbon atoms to the support gas plasma.

Once ignited, the source can operate with low input power 10-20 W. For a given source geometry and $B_z(r,\theta)$ profile, the discharge power is an essential parameter for the processes that lead to cluster formation within the plasma and the sooting of HC walls. Molecular gases and noble gas mixtures with $N_2$ have shown that one step clustering i.e., $C + C \rightarrow C_2$ and $N_1 + N_2 \rightarrow N_3$, may be favored but higher level of clustering $C_x + C_y \rightarrow C_z$ ($z \geq 3$) does not take place in the mixed gases' glow discharge plasma. The neon ignited plasma, on the other hand, has been observed to provide efficient HC sputtering that is conducive to the accumulation of soot layers on the walls and ionization of the sputtered species $C_m$. Sooting of the walls implies a dynamic plasma regenerative set-up. The interaction of the plasma and the soot generates bond breaking and re-bonding sequences that lead to cluster formation.

## 5. Conclusion

A continuous carbon cluster beam source has been described that relies on the clustering mechanisms initiated within the hollow cathode. It can be used to deliver a desired cluster beam for various experiments. Tunability of the source for any particular cluster depends upon the source geometry, the state of the HC wall sooting and the discharge power. It is a compact source with cylindrical dimensions 20 cm length 15 cm diameter. Six permanent bar magnets provide the cusp field that has been specially tailored with mild steel rings and plates to generate the desired $B_z(r,\theta)$ field contours.